\documentclass[12pt,preprint]{aastex}
\newcommand{\Msun}{~M_\odot}

\newcommand{\kms}{\rm ~km~s^{-1}}

\newcommand{\ergs}{\rm ~erg~s^{-1}}

\newcommand{\ml}{~\Msun ~\rm yr^{-1}}

\begin{document}

\title{X-RAYS FROM SUPERNOVA SHOCKS IN  DENSE MASS LOSS}
\author{Roger A. Chevalier and Christopher M. Irwin}
\author{}
\affil{Department of Astronomy, University of Virginia, P.O. Box 400325, \\
Charlottesville, VA 22904-4325; rac5x@virginia.edu}

\begin{abstract}
Type IIn and related supernovae show evidence for an interaction with a dense circumstellar medium
that produces most of the supernova luminosity.
X-ray emission from shock heated gas is crucial for the energetics of the interaction and
can provide diagnostics on the shock interaction.
Provided that the shock is at an optical depth $\tau_w\la c/v_s$ in the wind, where $c$ is the speed
of light and $v_s$ is the shock velocity, a viscous shock is expected that heats the gas to
a high temperature.
For $\tau_w\ga 1$, the shock wave is in the cooling regime; inverse Compton cooling dominates bremsstrahlung at higher densities and shock velocities.
Although $\tau_w\ga 1$, the optical depth through the emission zone is $\la 1$ so that inverse Compton effects do not give rise to significant X-ray emission.
The electrons may not reach energy equipartition with the protons at higher shock velocities.
As X-rays move out through the cool wind, the higher energy photons are lost to Compton degradation.
If bremsstrahlung dominates the cooling and Compton losses are small, the energetic radiation
can completely photoionize the preshock gas.
However, inverse Compton cooling in the hot region and Compton degradation in the wind  
reduce the ionizing flux, so that complete photoionization is not obtained and
photoabsorption by the wind further reduces the escaping
X-ray flux.
We conjecture that the combination of these effects led to the low observed X-ray flux
from the optically luminous SN 2006gy.
\end{abstract}


\section{INTRODUCTION}

There is increasing evidence for supernova shock waves propagating in dense, 
optically thick, mass loss regions.
Type IIn supernovae, which have narrow lines of hydrogen and other species in their
spectra \citep{schlegel90}, typically have  light curves where 
circumstellar interaction provides the power \citep[e.g.,][]{chugai92}.
The lines are indicative of continuing circumstellar interaction and the luminosity
implies a high circumstellar density.
Narrow H lines with broad, symmetric wings observed at early times can be
interpreted as the result of electron scattering in a medium with optical
depth of $3-4$ \citep{chugai01}.
Type IIn spectral features
have been observed in number of highly luminous supernovae,
including SN 2006gy \citep{smith07,ofek07}.
SN 2006gy, which was especially luminous in the optical,
 had an X-ray luminosity (or upper limit) that was orders
of magnitude below its optical luminosity \citep{smith07,ofek07}; the question
arises of whether the lack of a high X-ray luminosity 
is consistent with strong circumstellar
interaction \citep{katz11}.

There is the potential for X-ray emission provided that there is a viscous shock
wave that heats the gas in the circumstellar interaction.
A viscous shock front is expected to form in a dense
wind provided the optical depth in the external medium $\tau\la c/v_{s}$, where $v_{s}$ is the
shock velocity \citep{ofek10,chevalier11,nakar10}.
At larger optical depths the shock wave is mediated by radiation provided that
the ratio of radiation to matter pressure is $>4.45$ in the downstream region \citep{weaver76},
which is the case for the shocks considered here.
There is thus the potential for hot gas and its emission  from shock waves in
moderately optically thick regions \citep{katz11}.
Here we consider the X-ray emission from shocks in dense circumstellar regions,
concentrating on the optically thick case.
Any emitted X-ray emission has the possibility of being scattered or absorbed, 
changing both the supernova surroundings and the escaping X-ray radiation.
An understanding of how the shock power is eventually radiated is crucial for
understanding multiwavelength observations of these events.

We discuss the shock structure and X-ray emission in Section 2
and the implications for observations in Section 3.

\section{SHOCK PROPERTIES AND X-RAY EMISSION}
\label{diff}

We assume that the mass loss can be described by a steady wind flow;
the actual mass loss is unlikely to be steady, but this case should illustrate
the basic physical situation.
If the mass loss at rate $\dot M$ is in a steady wind at a velocity $v_w$,
the density  $\rho_w=\dot M/4\pi r^2v_w\equiv Dr^{-2}$ can be specified by a  density parameter, $D_*$,  scaled to a $\dot M=10^{-2}\ml$ and $v_w=10\kms$ wind
so that $\rho_w=5.0\times 10^{16}D_* r^{-2}$ in cgs units.
The optical depth in the wind to the dense shell position
$R$ is $\tau_w=1.7\times 10^{16}kD_*/R$ where $k$ is the opacity $\kappa$
in units of 0.34 cm$^2$ g$^{-1}$, appropriate for electron scattering with $n_{He}/n_H=0.1$.
If the expansion of $R$ can be expressed $R\sim t^m$ where $t$ is the
age \citep{chevalier03,chevalier11}, we have $v_s=mR/t$ so that 
\begin{equation}
D_*=0.06 v_{s4}\tau_wk^{-1}(t/10{\rm~day}),
\label{tau}
\end{equation}
where $v_{s4}$ is the expansion velocity of the shell $v_s$ in units of
$10^4\kms$ and $m=0.8$ has been assumed.
This relation with $\tau_w=1$ is shown in Fig.\ 1, which also shows the line
$\tau_w=c/v_{s}$ above which a viscous shock does not form.
In the figure, and in the following, we assume $k=1$.

\subsection{Radiation from the Shocked Region}

To give an estimate of the shock velocity, we note that for the supernova
model used by \cite{chevalier11} ($\rho\propto r^{-7}$ outer density profile), 
the velocity of gas at the reverse shock
wave is $6.3\times 10^3  E_{51}^{0.2}  M_{e1}^{-0.2}  D_*^{-0.2}
(t/10{\rm~ day})^{-0.2}\kms$,
where $E_{51}$ is the energy in units of $10^{51}$ ergs and $M_{e1}$ is the ejecta
mass in units of $10\Msun$.
The velocity $v$ of the interaction shell is also the shock velocity $v_{s}$ if the velocity
of matter immediately ahead of the shock is negligible.
There are two reasons that $v_{s}$ might be less than $v$.
One is that the presupernova wind has a significant velocity.
Wind velocities deduced from narrow H$\alpha$ line features
in Type IIn supernovae
are generally in the range $100-1000\kms$ \citep[Table 5 of][]{kiewe10}.
Second, there may be radiative acceleration of the mass loss gas.

To estimate the acceleration, we assume that the forward shock velocity is 
$\sim v$ and that the shock wave cools rapidly (compared to the age).
These assumptions should give the maximum luminosity and thus
the maximal effect of radiative acceleration.
The luminosity in this case is $L_c=2\pi R^2\rho_{w0}  v^3$, where $\rho_{w0}$
is the wind density immediately ahead of the shock wave.
The acceleration is $a_{rad}=L\kappa/4\pi R^2c$, where $c$ is the speed
of light, so that the gas is accelerated to
a velocity
\begin{equation}
v_{rad}=a_{rad}t\approx {\rho_{w0}v^2 mR\kappa \over 2c}.
\end{equation}
If we normalize to the shock velocity,
\begin{equation}
\frac{v_{rad}}{ v}=\frac{mDv\kappa}{2Rc}=\frac{m}{2}\frac{v}{c}\tau_w.
\end{equation}
The result is that $v_{rad}\sim v$ if $\tau_w\sim c/v$.
The formation of a viscous shock requires that $\tau_w\la c/v_s$, so we
expect significant acceleration near the time of shock breakout, if it occurs
in the wind region, and declining acceleration at lower optical depths.
The breakout radiation from a hot shell of radiation can also accelerate
the wind gas, but the luminosity of this event is comparable to the
luminosity that might be generated by the continuing shock interaction
\citep{chevalier11}.
\cite{katz11} obtain a similar result for the acceleration by breakout radiation and note
that the shock is collisionless.
Since the acceleration is important only at fairly large optical depth
and we will find that X-rays do not escape at those depths, we neglect
the acceleration here.    

The  cooling processes for the postshock hot gas 
are expected to be
bremsstrahlung and  inverse Compton cooling.
At the high densities of interest here, the dominant cooling emission is from the forward
shock region.
X-ray emission from supernovae is often discussed in terms the softer reverse
shock emission \citep{chevalier03}.
However, as the density increases, the reverse shock emission becomes limited by rapid cooling at
a lower circumstellar density than at the forward shock.
When both the forward and reverse shocks are in the radiative regime, the ratio
of forward to reverse shock power is $2(n-3)^2/(n-4)$, where $n$ is the
supernova density power law index \citep{chevalier03}.
For $n=7$, the ratio is 11.
In addition, the dense shell that forms between the shock fronts can absorb X-ray
emission produced at the reverse shock.
In the noncooling regime, the bremsstrahlung emission from the forward shock region is \citep{chevalier03}
\begin{equation}
L_b=3\times 10^{45} D_*^2(t/10{\rm~ day})^{-1}\ergs.
\end{equation}
In the cooling regime, the luminosity is
\begin{equation}
L_c=3.1\times 10^{44} D_* v_{s4}^3\ergs,
\end{equation}
where $v_{s4}$ is the shock velocity in units of $10^4\kms$.
The actual luminosity cannot be greater than that in the cooling case.
The cooling condition is that $L_b>L_c$, or
\begin{equation}
D_*>0.1 v_{s4}^3 (t/10{\rm~ day}).
\label{cool}
\end{equation}
The cooling condition is shown in Fig.\ 1; it is close to where the medium is optically thick.

Since the two sources of luminosity for inverse Compton cooling (shock breakout emission
and emission from continuing interaction) are comparable at early times and the
continuing interaction eventually can dominate,  
here we consider only the interaction luminosity source.
If the forward shock is cooling,  the
postshock luminosity is $L_c=2\pi R^2 \rho_0 v_s^3$.
We assume that a fraction $f$ of this luminosity is reradiated as approximately
blackbody thermal radiation and take $f=0.5$ as a reference value; backscattering
of the radiation can cause deviations from this value.
In the optically thick regime ($\tau_w>1$), the energy density in radiation is thus
\begin{equation}
u_{rad}\approx \frac{fL_c}{4\pi r^2 c}\tau_w.
\label{urad}
\end{equation}
The inverse Compton energy loss per unit volume is
\begin{equation}
\Lambda_C=4u_{rad}cn_e\sigma_T\frac{k_BT_e}{m_ec^2},
\label{lam}
\end{equation}
where $n_e$ is the electron density in the hot gas, $\sigma_T$ is the Thomson cross section, 
$k_B$ is Boltzmann's constant, $m_e$ is the electron mass, and 
$T_e$ is the electron temperature.
In the radiative, optically thick regime,
the ratio of the bremsstrahlung cooling rate to the Compton cooling rate is
\begin{equation}
\frac{\Lambda_{br}}{\Lambda_C}\approx 0.7\left(\frac{f}{0.5}\right)^{-1}v_{s4}^{-4} \tau_w^{-1},
\end{equation}
where equation (3-56) of \cite{spitzer78} was used for $\Lambda_{br}$.
The boundary between the cooling mechanisms is shown in Fig.\ 1.

Moving to the non-cooling case, we note that $L_c\propto D$ in the cooling case, but
the luminosity of the shell $L=V\Lambda_{br}\propto D^2$ in the noncooling case,
where $V$ is the emitting volume of the hot gas.
We again assume that the supernova luminosity is primarily from circumstellar
interaction so that the photospheric luminosity is $fV\Lambda_{br}$.
More specifically, we find
\begin{equation}
\frac{\Lambda_{br}}{\Lambda_C}\approx \frac{1}{4 f\tau_{0}}\left(\frac{kT}{m_ec^2}\right)^{-1}
=2.1\left(\frac{f}{0.5}\right)^{-1}v_{s4}^{-2}\tau_{0}^{-1}.
\end{equation}
The electron scattering optical depth through the hot shell, $\tau_{0}$, is the same as that
through the preshock wind gas, $\tau_w$, provided H and He are ionized and dominate the abundances,
and the shell gas is noncooling.
The value of $\tau_{0}$ drops below unity in the noncooling regime.
Inverse Compton cooling is important  at higher shock velocities and  wind
densities (Fig.\ 1).
This contrasts with the case where the radiation field is due to the
supernova, independent of the circumstellar interaction, when
bremsstrahlung emission becomes
a more important coolant at high density \citep{fransson82}.
At the higher densities in our case, inverse Compton again becomes important due to
the strong radiation field created by the interaction.
As indicated in Fig.\ 1, the condition that the shock front be cooling merges with the
transition between inverse Compton and bremsstrahlung cooling at higher shock velocities.
The reason for the absence of a region in which the shock is non-cooling with cooling
dominated by inverse Compton is our assumption that the supernova luminosity
is produced by the shock interaction.

If the nuclei and electrons do not rapidly achieve equilibrium
in the shock transition, the nuclei are heated to $T_p=
2.9\times 10^9v_{s4}^2$ K in the shock front.
The balance of electron cooling by inverse Compton losses with heating by Coulomb
collisions leads to a temperature $T_e\approx 7.1\times 10^8 \epsilon_{\gamma}^{-2/5}$ K,
where $\epsilon_{\gamma}$ is the fraction of the postshock energy density in
radiation and a Coulomb logarithm of 30 is assumed \citep{katz11}; $T_e$ may be higher
if there is collisionless heating of electrons.
From equation (\ref{urad}), we have $\epsilon_{\gamma}\approx (\tau_w v_s/c)$, so that
when the viscous shock first forms $\epsilon_{\gamma}\sim 1$, but it declines thereafter.
The electrons are heated to their equilibrium value ($1.4\times 10^9v_{s4}^2$ K)
when $\tau_w\approx 5.7v_{s4}^{-1}$.
Nonequilibrium is important only for high shock velocities (Fig.\ 1).
The criterion for equilibrium when the cooling is dominated by bremsstrahlung at lower densities is the same as that discussed by
\citet[][equation 12]{fransson82} and is shown in Fig.\ 1.

For $\tau_w\la 1$, bremsstrahlung is expected at close to the postshock temperature
and, since the emission goes in toward the dense photosphere as well as outward, with
a luminosity approximately equal to that from the photosphere.
As the optical depth to the emission region increases,  there is more of a chance of an
outward going photon being scattered in toward the photosphere, so the hard X-ray luminosity
declines relative to the photospheric emission.
In the  cooling case, the electron scattering optical depth through the hot, shocked region is
$\tau_0=n_e \sigma_Td$, where  $d=(v_s/4)t_{cool}$ and $t_{cool}=(3/2)nk_BT/\Lambda$.
If inverse Compton dominates the cooling and $\tau_w>1$, $\Lambda$ is given by equations 
(\ref{urad}) and (\ref{lam}), and we have 
\begin{equation}
\tau_0=\frac{3}{4f\mu \tau_w}\frac{m_e}{m_p}\left(\frac{c}{v_s}\right)^2
\approx\frac{0.6}{f\tau_w v_{s4}^2},
\end{equation}
where $\mu$ is the mean particle weight divided by the proton mass $m_p$.
We thus find that $\tau_0\la 1$, so that a typical outgoing photon from the dense shell scatters at
most once in the hot gas; since the electron energies are $\la m_ec^2$, the photon energy
is changed by a factor $<2$.
For $\tau_w>1$, photons can be scattered back through the hot region, but production of
photons up to X-ray energies is not expected in this way because the ingoing photons are
absorbed by the dense shell and ejecta.
We thus expect that, in the regime where inverse Compton cooling dominates, the X-ray emission
declines relative to emission at lower photon energies.
This is also true when bremsstrahlung cooling dominates and $\tau_w>1$ because
initially outgoing photons can be scattered back across the emission region and absorbed in
the dense shell.

\subsection{Photon Interaction with the Preshock Wind}

The X-ray radiation from the hot shell must escape through the cooler unshocked envelope
of circumstellar gas.
Both scattering and absorption of the photons can be important.
Comptonization affects the escape of photons at the high energy end.
The maximum energy is $\epsilon_{max}\approx m_ec^2/\tau_{es}^2=511/\tau_{es}^2$
keV, where the electron scattering optical depth $\tau_{es}=\tau_w$ for our assumptions.
This result is due to the fact that electron recoil gives a change of photon wavelength
$\sim h/m_ec$ for each scattering, where $h$ is Planck's constant, 
and the number of scatterings is $\sim \tau_{es}^2$.
Detailed calculations show that there is not a sharp cut-off at the energy, but
the spectrum becomes steep \citep[e.g., Fig. 15 of][]{kylafis82}.
In Section 1, we noted that a viscous shock could form at $\tau_{w}\approx c/v_s=30v_{s4}^{-1}$ so that $\epsilon_{max}$ can be as small as $0.57 v_{s4}^{2}$ keV. 
The emission from the hot gas 
goes into heating and ionization of the preshock gas until the shock wave reaches moderate
optical depth.

Photoionization of the preshock medium is important for the absorption of X-rays
from the shocked region.
If the preshock medium is completely ionized, we expect relatively little absorption, while
incomplete ionization leads to substantial absorption for our parameters.
In considering the photoionization of the preshock gas, we will be assuming that the
photoionization is determined by the current X-ray luminosity, i.e. that steady state
conditions apply.
This requires that the recombination time be less than the age, which is generally
true for the high densities considered here.

If the medium is optically thin, photoionization is  related to the ionization parameter
$\xi=L/nr^2$ in cgs units, where $n$ is the density and $r$ is the distance from
the luminosity source.
For an $r^{-2}$ density distribution, 
the $\xi$ parameter is independent of radius.  
If we are in the cooling regime and the ionizing luminosity is $\beta L_c$, 
the ionization parameter is $\xi\approx 1\times 10^4 \beta v_{s4}^3$,
independent of $D$.
For lower values of $D$ in the non-cooling regime, we have
$\xi\propto Dt^{-1}$.
The ionization parameter is highest in the high density regime of interest here
and drops at lower densities in the noncooling regime.
Photoionization calculations have been previously carried out and we briefly 
summarize results for an emitting gas temperature of 10 keV, or
$\sim 10^8$ K.
For a value $\xi\sim 10^4$, the medium is completely ionized 
\citep{tarter69,hatchett76,kallman82}.
The elements C, N, and O are completely ionized for $\xi\ga 100$; ionization of
the heavier elements (S and Fe) requires $\xi >10^3$ \citep{hatchett76}.

To extend these results to a hotter radiation field, we used the photoionization code
CLOUDY \citep{fer98}.
We started by using a radiation field set by a $T=10^8$ K bremsstrahlung spectrum; the
results were consistent with those described above.
We went to higher bremsstrahlung temperatures to allow for higher shock velocities;
at $v_{s4}=1$, the postshock temperature is $\sim 10^9$ K.
At $\xi=1000$, 
changing the bremsstrahlung temperature does not have much effect on
ionization structure or $T_e$.  For $\xi\la 1000$, a higher
temperature luminosity leads to a lower $T_e$ and less
ionization.  But for $\xi\ge 1000$, a higher temperature luminosity produces a
higher $T_e$ and more ionization.
For example, at $T=10^8$ K, the CNO elements are completely ionized at $\xi=100$, S
becomes ionized at $\xi\sim 1000$, and everything is ionized at $\xi\sim 10^4$. 
However, for $T=10^9$ K, CNO do not become ionized until $\xi\sim 500$, S still
becomes ionized at $\xi=1000$, but everything is ionized at $\xi\sim 5000$.
These results can be understood since the ionization potential is $\propto Z^2$,
where $Z$ is the atomic number.
The higher temperature emission has higher energy photons, which are less efficient
at ionizing atoms with lower ionization potentials (compared to the photon energy),
but are more effective at ionizing atoms with high ionization potentials.

For $\tau_w\sim 1$, we expect that $\beta\sim 0.5$, and the value of $\xi$ mainly
depends on $v_s$.
For $v_{s4}\ge1$, the gas is completely ionized, while for $v_{s4}=0.5$ only
partial ionization is likely.
Another issue is whether the preshock medium is optically thick
in the photoionization continua, and the incident radiation
field is depleted in crossing the circumstellar gas.
Using the results of \citet{TS69}, we find that at the higher
temperatures of interest ($\sim 10^9$ K) for the ionizing radiation,  photon depletion is negligible and ionization remains nearly constant with radius.
At lower temperatures for the radiation ($\sim 10^8$ K), there is a transition to
the Stromgren regime.

The situation changes if the wind is moderately optically thick to electron scattering.
An effect of electron scattering is to reverse the motion of photons, so that the
energy density of ionizing photons is increased; the time spent by photons in a
region is increased by $\sim \tau_{w}$ (equation [\ref{urad}]), so that the rate of ionizations (and $\xi$) is increased
by  the same factor \citep[e.g.,][]{ross79}.
While this effect favors higher photoionization and the escape of X-rays, there are several effects that disfavor
high X-ray emission.
First, Compton degradation in the wind leads to the loss of the high energy photons, as discussed above;
emission above 2 keV is suppressed when $\tau_w\sim 16$.
Second, at moderate optical depths inverse Compton cooling tends to dominate 
bremsstrahlung (Fig.\ 1), so that X-ray emission is a smaller fraction of the shock power;
inverse Compton is more important at higher shock velocities.
In addition, initially outgoing X-ray photons have some chance of being scattered back and
absorbed by the dense shell.
Finally, at optical depths $\ga 10$, these two effects are likely to be larger than the
increase in the ionizing radiation field, so that the ionization parameter is decreased
and there is an increased chance of photoabsorption of the X-ray emission.
Photoabsorption is important at low X-ray energies while Compton degradation
is important at high X-ray energies.
An accurate calculation of the X-ray emission is complicated and beyond the current paper.
An analogous physical situation is X-ray emission from optically thick accretion
onto white dwarfs \citep{kylafis82}.  
The photon trapping limit $\tau_w=c/v_s$ here corresponds to an accretion rate at the
Eddington limit in the accretion case.
A difference is that the emission from the white dwarf surface can maintain complete
ionization of the preshock gas, which is not necessarily the case here.

\section{IMPLICATIONS FOR OBSERVATIONS}

Most of the X-ray observations of Type IIn supernovae in dense media are an age $>1$ yr and
sometimes much greater, so that the objects do not fall in the parameter space
shown in Fig. 1.
However, SN 2006gy, with an estimated $D_*\approx 10$ \citep{smithmccray07,chevalier11}, was observed
with {\it Chandra} on 2006 Nov 14, at an age of $3-4$ months \citep{smith07,ofek07}.
\cite{smith07} inferred a detection, with all counts at energies $<2$ keV and an unabsorbed
$0.5-2$ keV luminosity of $1.65\times 10^{39}\ergs$ assuming $T=1$ keV.
From the same data, \cite{ofek07} inferred a nondetection with an upper limit of $1.6\times 10^{40}\ergs$, assuming
a photon index of 1.8.
In either case, the X-ray luminosity was much less than the observed photospheric luminosity of
$\sim 3\times 10^{44}\ergs$ \citep{smith10}.
In the model of \cite{chevalier11} for SN 2006gy, the shock wave radiation broke out in
the mass loss region, so the optical depth outside the shock was initially $c/v_s$.
The parameters for the X-ray observation give $D_*(t/10{\rm~ day})^{-1}\sim 1$, near the
$\tau_w=c/v_s$ line in Fig.\ 1.
\cite{smith10} estimate $v_s\approx 4000-5000\kms$ near the time of peak optical
luminosity so, as discussed in Section 2.2, Comptonization by itself could limit the
escaping photons to $<0.14$ keV, and the loss of ionizing radiation would allow photoabsorption.

Our discussion suggests the following sequence, as the density of 
surrounding mass loss increases.
At low density, the optical luminosity is dominated by radioactivity and shock
heating of the progenitor; X-ray emission from interaction is initially a small part of the luminosity,
although it might become a larger part at late times when other power sources fade.
As the optical depth in the wind approaches unity, interaction typically dominates the
power input and X-ray emission can be a signficant part of the power.
At higher densities, the X-ray emission falls relative to the optical luminosity because
of inverse Compton cooling of the shocked region, Compton degradation in the wind, and photoabsorption.

\acknowledgments
We thank   Claes Fransson  for discussions and the referee for useful comments.
This research was supported in part by NSF grant  AST-0807727.


\clearpage

\begin{figure}[!hbtp]   
\epsscale{1.0}
\plotone{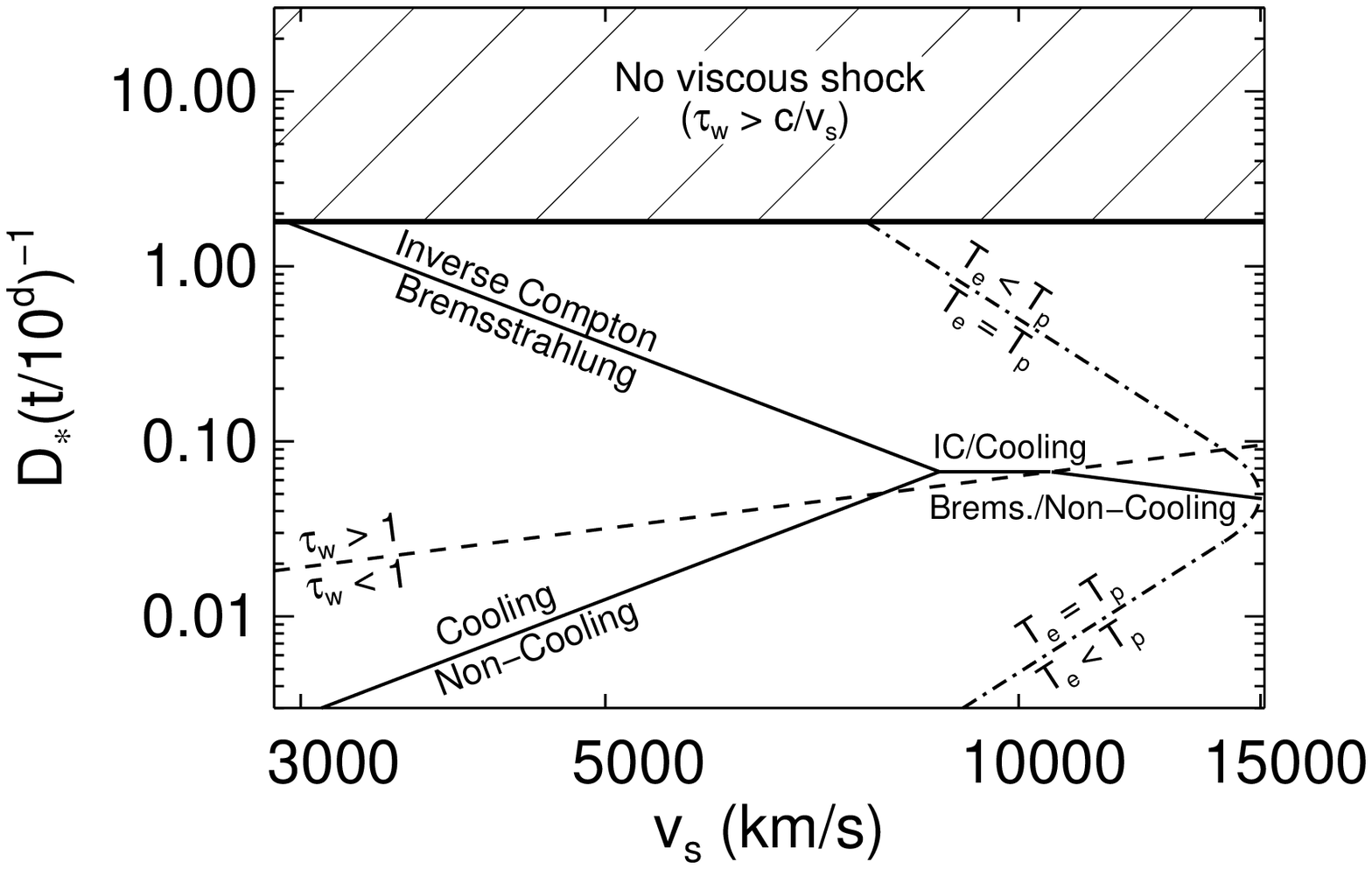}
\caption{Regimes of shock structure in terms of wind density/age  vs.\ shock velocity.
The lines in the figure mark the limits of $a$) presence of viscous shock (thick solid line), $b$) optical depth unity (dashed),
$c$) rapid cooling (solid), $d$) cooling by inverse Compton vs.\ bremsstrahlung (solid), and $e$) electron-proton equilibration (dash-dot).  
}
\end{figure}

\end{document}